\documentstyle[12pt]{article}
\def\d{\partial}

\def\kaa{K\"ahler}
\def\F{{\cal F}}
\def\A{{\cal A}}

\def\bop#1{\setbox0=\hbox{$#1M$}\mkern1.5mu
        \vbox{\hrule height0pt depth.04\ht0
        \hbox{\vrule width.04\ht0 height.9\ht0 \kern.9\ht0
        \vrule width.04\ht0}\hrule height.04\ht0}\mkern1.5mu}

\begin{document}

\newcommand{\beq}{\begin{equation}}
\newcommand{\eeq}[1]{\label{#1}\end{equation}}
\newcommand{\ber}{\begin{eqnarray}}
\newcommand{\eer}[1]{\label{#1}\end{eqnarray}}
\begin{center} March, 1995		\hfill    ITP-SB-95-5\\
\hfill    USITP-95-4\\
\hfill    hep-th/9503012\\

\vskip .3in

{\large \bf  A note on the Seiberg-Witten solution of \\ \ \\ N=2 Super
Yang-Mills
Theory}
\vskip .3in

{\bf Ulf Lindstr\"om} \footnotemark \\

\footnotetext{e-mail address: ul@vana.physto.se}

{\em  Institute of Theoretical Physics \\ University of Stockholm \\ Box 6730
\\
S-113 85 Stockholm SWEDEN}\\

\vskip .15in

{\bf Martin Ro\v cek} \footnotemark \\

\footnotetext{e-mail address: rocek@insti.physics.sunysb.edu}

{\em Institute for Theoretical Physics \\ State University of New York at Stony
Brook \\ Stony Brook, NY 11794-3840 USA}\\
\vskip .1in
\end{center}
\vskip .4in
\begin{center} {\bf ABSTRACT } \end{center}
\begin{quotation}\noindent We examine the Seiberg-Witten treatment of $N=2$
super
Yang-Mills theory, and note that in the strong coupling region of moduli space,
some massive particle excitations appear to have negative norm. We discuss the
significance of our observation.
\end{quotation}
\vfill
\eject
\def\baselinestretch{1.2}
\baselineskip 16 pt
\noindent

Recently, Seiberg and Witten solved the low energy limit of
$N=2$ super Yang-Mills theory with gauge group $SU(2)$ \cite{sw}.   Their
solution
is described by a single holomorphic function $\F$, which determines the
geometry
of the \kaa-vector multiplet \cite{kvm}, and is generated by one-loop and
instanton
effects. The vacuum of $N=2$ super Yang-Mills theory is described by a complex
parameter $u$ that describes the spontaneous symmetry breaking from
$SU(2)$ to $U(1)$. They explored in great detail the couplings of the
$U(1)$ gauge
multiplet to the low-lying excitations of the theory and uncovered a rich
structure
involving duality. They also briefly mentioned the couplings to some of the
massive
states.  In this note we push some of those ideas a little further, and find a
surprising result.

Seiberg and Witten compute $\F$ as a function of the chiral $N=2$
superfield $\A$
describing the unbroken $U(1)$ gauge group.  This function determines the low
energy physical parameters of the theory.  For example,
\beq
\tau\equiv \left\langle\F ''\right\rangle
=\frac{\theta}{2\pi}+i\frac{4\pi}{g^2}
\eeq{tau} gives the gauge coupling $g^2$ in terms of $Im (\left\langle \F
''\right\rangle)$.  This is positive because $\tau$ can be interpreted as the
modular parameter of a family of tori, and takes its values in the upper
half-plane.

Seiberg and Witten also note that new phenomena occur when
\beq Im (\left\langle\frac{\F '}\A \right\rangle) = 0\ .
\eeq{imo} They suggest, and it has been verified numerically, that this
occurs on a
curve (topologically a circle) in the space of vacua \cite{imo}. Outside of
that
curve, the semiclassically derived spectrum holds, and $Im
(\left\langle\frac{\F
'}\A \right\rangle) > 0$; inside, states may disappear and/or appear, and $Im
(\left\langle\frac{\F '}\A \right\rangle) < 0$.

As they note, $SU(2)$ gauge invariance implies that
$\F(\A)\equiv\F(\sqrt{\A^i\A^i})$, where $\A^i$ is the triplet of gauge
superfields
of the full group $SU(2)$.  Then instead of a complex constant $\tau$, one
finds a
$3\times 3$ matrix of couplings
\beq
\tau_{ij}\equiv \left\langle\frac{\d^2}{\d
\A^i\d\A^j}\F(\sqrt{\A^k\A^k})\right\rangle =
\left\langle\F '' \frac{\A^i\A^j}{\A^k\A^k} +\frac{\F '}{\sqrt{\A^k\A^k}}
(\delta^{ij}- \frac{\A^i\A^j}{\A^l\A^l})\right\rangle\ .
\eeq{tauij} What was apparently overlooked in \cite{sw} is the nice
decomposition
(\ref{tauij}) of $\tau_{ij}$ into color projection operators onto the
$U(1)$ and
the $SU(2)/U(1)$ generators.  The coefficient of the $U(1)$ fields is just
$\left\langle\F ''\right\rangle$ ({\it i.e.\/}, $\tau$), which, as noted in
\cite{sw}, leads to a positive kinetic energy. The coefficient of the coset
fields
is instead $\left\langle\frac{\F '}\A \right\rangle$ (in the notation of
\cite{sw},
this is also $a_D/a$).  As noted above, the imaginary part of this vanishes on
a
curve in the space of vacua; on that curve, we expect the corresponding gauge
fields to become nonpropagating, auxiliary degrees of freedom
\cite{glr}. Inside the curve, the would-be particles have negative kinetic
terms.
This signals a breakdown of the formalism for the coupling to the charged
massive
$W$ vector gauge multiplets. Presumably, it is a signal that the $W$ particles
have disappeared from the spectrum (as anticipated in \cite{sw}); however, from
the effective action, we do not know how to prove that negative norm states are
safely nonexistent rather than a signal of a genuine breakdown of the theory.

Duality transformations cannot illuminate the difficulty, as they map the
inside of
the curve into itself, and thus do not mix the semiclassical region with this
strongly coupled region.

We end with a few comments: Our results do not in any way challenge the
validity of
the solution of \cite{sw} for the low energy excitations; it seems to be an
added
bonus that in the semiclassical region, the solution also describes the
coupling to
the massive $W$'s correctly, but there is no reason for this to hold in the
strong
coupling domain, as the masses are above the cutoff in the Wilson effective
action.
The question that we have stumbled on in the context of $N=2$ super Yang-Mills
theory is quite general: when do pathologies of the effective action reflect a
breakdown of the theory, and when do they represent a signal that the states
described by the effective action have disappeared?

\vskip .3in \noindent {\bf Acknowledgements} \vskip .2in \noindent We would
like to
thank J. de Boer, D. Jatkar, B. Peeters, K. Skenderis, G. Sterman, P. van
Nieuwenhuizen, W. Weisberger, and E. Witten for discussions. The work of MR was
supported in part by  NSF grant No.\ PHY 9309888. The work of UL was
supported in
part by NFR grant No.\ F-AA/FU 04038-312 and by NorfA grant No.\ 94.35.049-O.


\vskip .3in

\begin{thebibliography}{6666}

\newcommand{\np}{Nucl. Phys. }
\newcommand{\pr}{Phys. Rev. }
\newcommand{\prl}{Phys. Rev. Lett. }
\newcommand{\cmp}{Commun. Math. Phys. }
\newcommand{\pl}{Phys. Lett. }

\bibitem{sw} N. Seiberg and E. Witten, \np {\bf B426} (1994) 19.

\bibitem{kvm} S. J. Gates, Jr., \np {\bf 238} (1984) 349;\\ B. de Wit, P. G.
Lauwers, R. Philippe, S.-Q. Su and A. Van Proeyen,
\pl {\bf 134B} (1984) 37;\\ M. G\"unaydin, G. Sierra and P. K. Townsend,
\np {\bf
B242} (1984) 244.

\bibitem{imo} J. de Boer, D. Jatkar, C. Skenderis, B. Peeters, unpublished.\\
Argyres and Faraggi, unpublished (E. Witten, private communication).

\bibitem{glr} J. Grundberg, U. Lindstr\"om, and M. Ro\v cek, in
preparation.\\ {\em
See also} B. Julia and J. F. Luciani, \pl {\bf 90B} (1980) 270.\\ C. M. Hull,
A.
Karlhede, U. Lindstr\"om and M. Ro\v cek, \np {\bf B266}  (1986) 1.


\end{thebibliography}
\end{document}